\documentstyle[12pt]{article}

\def\GeV{\,{\rm GeV}}

\def\Mpc{\,{\rm Mpc}}

\def\cmm2{{\,\rm cm^{-2}}}
\def\cm2{{\,{\rm cm}^2}}
\def\cmm3{{\,{\rm cm}^{-3}}}
\def\gcmm3{{\,{\rm g\,cm^{-3}}}}

\def\mpl{{m_{\rm Pl}}}

\def\la{\mathrel{\mathpalette\fun <}}
\def\ga{\mathrel{\mathpalette\fun >}}
\def\fun#1#2{\lower3.6pt\vbox{\baselineskip0pt\lineskip.9pt
  \ialign{$\mathsurround=0pt#1\hfil##\hfil$\crcr#2\crcr\sim\crcr}}}

\begin{document}
\pagestyle{empty}

\begin{center}
\rightline{FERMILAB--Pub--93/026-A}
\rightline{astro-ph/9302***}
\rightline{submitted to {\it Physical Review D}}

\vspace{.2in}
{\Large\bf On the Production of Scalar and Tensor Perturbations
in Inflationary Models}\\

\vspace{.2in}
Michael S. Turner \\

{\it Departments of Physics and of Astronomy \& Astrophysics\\
Enrico Fermi Institute, The University of Chicago, Chicago, IL  60637-1433}\\

{\it NASA/Fermilab Astrophysics Center,
Fermi National Accelerator Laboratory, Batavia, IL  60510-0500}\\

\end{center}

\vspace{.15in}

\centerline{\bf ABSTRACT}

\noindent
Scalar (density) and tensor (gravity-wave) perturbations
provide the basis for the fundamental observable consequences
of inflation, including CBR anisotropy and the fluctuations
that seed structure formation.  In general, these perturbations
are nearly scale invariant (Harrison-Zel'dovich spectrum),
though slight deviation from scale invariance (``tilt'') can have significant
consequences for both CBR anisotropy and structure formation.
In particular, a slightly tilted spectrum of scalar
perturbations may improve the agreement of the cold dark
matter scenario with the present observational data.
The amplitude and spectrum of the scalar and tensor
perturbations depend upon the
shape of the inflationary potential in the small interval
where the scalar field responsible for inflation
was between about 46 and 54 e-folds before the end of inflation.
By expanding the inflationary potential in a Taylor series in this interval
we show that the amplitude of the perturbations and
the power-law slope of their spectra can be expressed in
terms of the value of the potential 50 e-folds before
the end of inflation, $V_{50}$, the steepness
of the potential, $x_{50} \equiv \mpl V_{50}^\prime /V_{50}$,
and the rate of change of the steepness, $x_{50}^\prime$ (prime
denotes derivative with respect to the scalar field).
In addition, the power-law index of the cosmic-scale
factor at this time, $q_{50}\equiv [d\ln R/d\ln t]_{50}
\simeq 16\pi /x_{50}^2$.  (Formally, our results for
the perturbation amplitudes and spectral indices are
accurate to lowest order in the deviation from scale invariance.)
In general, the deviation from scale invariance is such to enhance
fluctuations on large scales, and is
only significant for steep potentials, large $x_{50}$, or
potentials with rapidly changing steepness,
large $x_{50}^\prime$.  In the latter
case, only the spectrum of scalar perturbations
is significantly tilted.  Steep potentials are
characterized by large tensor-mode contribution to
the quadrupole CBR temperature anisotropy, similar
tilt in both scalar and tensor
perturbations, and slower expansion rate, i.e., smaller $q_{50}$.
Measurements of the amplitude and tilt of the
scalar and tensor perturbations over determine $V_{50}$, $x_{50}$,
and $x_{50}^\prime$, and can in principle
be used to infer these quantities as well as
testing the inflationary hypothesis.  Our formalism has
its limitations; it is not applicable to
potentials with unusual features in the region
that affects astrophysical scales.

\pagestyle{plain}
\setcounter{page}{1}
\newpage

\section {Introduction}

In inflationary Universe models \cite{guth,inflation}
scalar (density) and tensor (gravity-wave)
metric perturbations arise due to de Sitter-space
produced quantum fluctuations.
The production of both density perturbations
\cite{scalar} and gravity-wave perturbations \cite{tensor}
have been well studied and are by now well understood.
Very roughly, the quantum fluctuations on a given length scale become
classical metric perturbations when that scale (Fourier mode)
crosses outside the Hubble radius during inflation, that is,
when $\lambda_{\rm phys} = R\lambda \sim H^{-1}$.  Here $R$ is
the cosmic-scale factor, $\lambda$ is the comoving wavelength
of the Fourier mode, and $H$ is the Hubble parameter.
The scales of astrophysical interest, say
from galaxy-size perturbations of $1\Mpc$ to the present
Hubble scale of $10^4\Mpc$, cross outside the Hubble radius
about 50 or so e-foldings in the scale
factor before the end of inflation, over an a span of about
$\ln (10^4)\sim 8$ e-folds.  In the post-inflationary
Universe scalar-mode perturbations re-enter the Hubble radius
with an amplitude that is approximately scale invariant:
$(\delta\rho /\rho)_{\rm HOR}\sim [V^{3/2}/V^\prime \mpl^3]
\lambda_{\Mpc}^{\ \ \,\alpha_S}$,
$|\alpha_S| \ll 1$; $\alpha_S = 0$ corresponds to the
Harrison-Zel'dovich scale-invariant spectrum \cite{hz}.
Likewise, the tensor-mode perturbations re-enter the Hubble radius
with a dimensionless amplitude (gravitational-wave strain)
that is approximately scale invariant:  $h_{\rm GW} \sim
[V^{1/2}/\mpl^2]\lambda_{\Mpc}^{\ \ \,\alpha_T}$,
$|\alpha_T| \ll 1$.  Here $V(\phi )$ is the inflationary
potential, $\mpl = 1.22\times 10^{19}\GeV$ is the Planck
mass, and $\lambda_{\Mpc} =\lambda /\Mpc$.  The power-law
indices $\alpha_S$ and $\alpha_T$ are
related to the those frequently used
to characterize the power spectra of the perturbations:
$n = 1 - 2\alpha_S$ and $n_T=-2\alpha_T$.

The scalar perturbations provide the primeval density
fluctuations that seed structure formation in
cold dark matter scenarios, and so their amplitude
and spectrum are of crucial importance.
Both scalar- and tensor-mode perturbations can lead to
temperature fluctuations in the cosmic background radiation
(CBR), as briefly summarized in the Appendix.
On angular scales much larger than a degree, the scale
subtended by the Hubble radius at decoupling, the scalar
and tensor contributions are inseparable; on smaller
angular scales the contribution of the tensor-mode
perturbations becomes subdominant and the angular
dependence of CBR anisotropy can in principle
be used to separate the scalar and tensor
contributions.  The detection of anisotropy in the CBR on angular
scales greater than about $10^\circ$ by the DMR instrument on COBE
\cite{dmr} spurred interest as to the mix of
scalar and tensor contribution to large-angle CBR anisotropy
\cite{krauss}, and what can be learned
about inflationary models \cite{davisetal}.
The purpose of this paper is to
relate the amplitude and spectrum of the scalar and
tensor perturbations to the shape of the inflationary potential.
Some of the issues, e.g., deviations from scale invariance
\cite{st} and the relative contributions of scalar and
tensor perturbations to the quadrupole anisotropy \cite{davisetal},
have been addressed elsewhere; in addition to extending
previous work in several important regards,
we have attempted to concisely and clearly
relate specific properties of the inflationary potential to
the potentially measurable features of the metric
perturbations \cite{extenddavis}.

In the next Section we briefly review slow-rollover inflation
and the production of metric perturbations; since all the
observable effects of these perturbations involve the shape
of the potential over an interval of only about 8 e-folds around
50 e-folds before the end of inflation, we expand
the potential about this point in terms of its value, $V_{50}$,
its steepness, $x_{50}\equiv [\mpl V^\prime /V]_{50}$,
and the change in its steepness, $x_{50}^\prime$ (prime
denotes derivation with respect to the scalar field).
We show that the amplitude and scale dependence of scalar and
tensor perturbations, quantified as $\alpha_S$ and
$\alpha_T$, are simply related to these quantities,
and further that the rate of growth of the cosmic-scale
factor around 50 e-folds before the end of inflation
is related to the steepness of the potential.  In Section III
we apply our formalism to four different types of inflationary
potentials, and draw some general conclusions.
The deviations from scale invariance tend to enhance
large-scale perturbations.  The models that
have significant deviation from scale invariance
involve either steep potentials or potentials
with rapidly changing steepness.
In the latter case, only the scalar perturbations
are tilted significantly.  In the case of steep potentials,
the scalar and tensor perturbations are tilted
by a similar amount.   The relative contributions
of the scalar and tensor perturbations to the quadrupole CBR
anisotropy is related to steepness of the potential (and
hence the deviation from scale invariance):  Large tensor contribution
implies significant deviation from scale invariance, and slower
expansion rate during inflation.  In Section IV we finish
with some concluding remarks.

\section{Inflationary Perturbations}

All viable models of inflation are of the
slow-rollover variety, or can be recast as
such \cite{inflation,allslow}.  In slow-rollover
inflation a scalar field
that is initially displaced from the minimum
of its potential rolls slowly to that minimum, and as it does
the cosmic-scale factor grows very rapidly; the Universe
is said to inflate.   Once the scalar field reaches the
minimum of the potential it oscillates about it, so that the large
potential energy has been converted into coherent
scalar-field oscillations, corresponding to
a condensate of nonrelativistic scalar particles.
The eventual decay of these particles into lighter
particle states and their subsequent thermalization
lead to the reheating of the Universe to a
temperature $T_{\rm RH} \simeq \sqrt{\Gamma \mpl}$,
where $\Gamma$ is the decay width of the
scalar particle \cite{reheat,allslow}.
Quantum fluctuations in the scalar field driving inflation
lead to scalar metric perturbations (referred to density or curvature
perturbations) \cite{scalar}, while quantum fluctuations
in the metric itself lead to tensor metric perturbations
(gravity waves) \cite{tensor}; isocurvature perturbations
can arise due to quantum fluctuations in other massless fields,
e.g., the axion field, if it exists \cite{isocurv}.

We assume that the scalar field driving inflation is
minimally coupled so that its stress-energy tensor
takes the canonical form,
\begin{equation}
T_{\mu\nu} = \partial_\mu\phi \partial_\nu\phi
-{\cal L}g_{\mu \nu};
\end{equation}
where the Lagrangian density of the scalar field
${\cal L} = {1\over 2}\partial_\mu\phi\partial^\mu\phi - V(\phi )$.
If we make the usual assumption that the scalar field
$\phi$ is spatially homogeneous, or at least so over a Hubble radius,
the stress-energy tensor takes the perfect-fluid form
with energy density, $\rho = {1\over 2}{\dot\phi}^2
+V(\phi )$, and isotropic pressure, $p={1\over 2}{\dot\phi}^2 -V(\phi )$.
The classical equations of motion
for $\phi$ can be obtained from the first law of thermodynamics,
$d(R^3\rho )=-pdR^3$, or by taking the four-divergence of $T^{\mu\nu}$:
\begin{equation}\label{eq:sr}
\ddot\phi + 3H\dot\phi + V^\prime (\phi ) = 0;
\end{equation}
the $\Gamma \dot\phi$ term responsible for reheating has
been omitted since we shall only be interested in the
slow-rollover phase.  In addition, there is the
Friedmann equation, which governs the expansion of the Universe,
\begin{equation}\label{eq:feq}
H^2 = {8\pi \over 3\mpl^2}\left( V(\phi ) + {1\over 2}
{\dot\phi}^2\right) \simeq {8\pi V(\phi )\over 3 \mpl^2};
\end{equation}
where we assume that the contribution of
all other forms of energy density, e.g., radiation and kinetic
energy of the scalar field, and
the curvature term ($k/R^2$) are negligible.
The justification for discussing inflation in the context
of a flat FRW model with a homogeneous scalar field
driving inflation are discussed at length in Ref.~\cite{inflation};
including the $\phi$ kinetic
term increases the righthand side of Eq. (\ref{eq:feq}) by a factor
of $(1+x^2/48\pi )$, a small correction for viable models.

Later in this Section and in the Appendix we will be more
precise about the amplitude of density perturbations and
gravitational waves; for now it will suffice
to give the characteristic amplitude of each:
\begin{eqnarray}\label{eq:scalartensor}
(\delta\rho /\rho)_{{\rm HOR},\lambda} & =  & c_S \left(
{V^{3/2}\over \mpl^3 V^\prime}\right)_1; \\
h_{{\rm HOR},\lambda} &  =  &
c_T \left({V^{1/2}\over \mpl^2}\right)_1 ;
\end{eqnarray}
where $(\delta\rho /\rho )_{{\rm HOR},\lambda}$ is the amplitude
of the density perturbation on the scale $\lambda$
when it crosses the Hubble radius during the post-inflation epoch,
$h_{{\rm HOR},\lambda}$ is the dimensionless amplitude
of the gravitational wave perturbation on the scale $\lambda$
when it crosses the Hubble radius, and $c_S$, $c_T$
are numerical constants of order unity.
Subscript 1 indicates that the quantity involving
the scalar potential is to be evaluated when the
scale in question crossed outside the
horizon during the inflationary era.

[Two small points; in Eq. (\ref{eq:scalartensor})
we got ahead of ourselves and used the slow-roll
approximation (see below)
to rewrite the fundamental expression, $(\delta \rho /\rho )_{{\rm HOR},
\lambda} \simeq (V/\mpl^2{\dot\phi})_1$
\cite{scalar}, in terms of the potential only.  While we shall
always mean ``cross outside'' or ``inside the Hubble radius,''
we will occasionally slip and say, ``cross outside''
or ``inside the horizon'' instead.]

Eqs. (\ref{eq:sr}-5) are the fundamental equations that govern
inflation and the production of metric perturbations.
It proves very useful to recast these equations using
the scalar field as the independent variable; we
then express the scalar and tensor perturbations
in terms of the value of the potential, its steepness,
and the rate of change of its steepness when the interesting
scales crossed outside the Hubble radius during inflation,
about 50 e-folds in scale factor before the end of inflation,
defined by
$$V_{50} \equiv V(\phi_{50}); \qquad
x_{50} \equiv {\mpl V^\prime (\phi_{50})\over
V(\phi_{50})}; \qquad x_{50}^\prime = {\mpl V^{\prime\prime}
(\phi_{50})\over V(\phi_{50})} - {\mpl [V^\prime (\phi_{50})]^2
\over V^2(\phi_{50})}.$$
And as we shall discuss, we will work to lowest order in
the deviations from scale invariance, $\alpha_S$ and $\alpha_T$,
which corresponds to order $x_{50}^2$, $\mpl x_{50}^\prime$.
Terms involving higher-order derivatives of the potential
lead to corrections that are higher-order in the deviation from scale
invariance.

To evaluate these three quantities 50 e-folds before
the end of inflation we must find the value of
the scalar field at this time.
During the inflationary phase the $\ddot\phi$ term is
negligible (the motion of $\phi$ is friction dominated),
and Eq. (\ref{eq:sr}) becomes
\begin{equation}\label{eq:sra}
\dot\phi \simeq {-V^\prime (\phi) \over 3H};
\end{equation}
this is known as the slow-roll approximation \cite{st}.
(The corrections to the slow-roll approximation
are ${\cal O}(\alpha_i)$ for the amplitude of
perturbations, and ${\cal O}(\alpha_i^2)$
for the power-law indices themselves.  There are models where
the slow-roll approximation cannot be used at all;
e.g., a potential where during the crucial 8 e-folds
the scalar field rolls uphill, ``powered'' by the
velocity it had when it hit the incline.)

The conditions that must be satisfied in order that
$\ddot\phi$ be negligible are:
\begin{eqnarray}\label{eq:condx}
|V^{\prime\prime}| & < & 9H^2 \simeq 24\pi V/\mpl^2; \\
|x| \equiv |V^\prime \mpl /V|  & <  & \sqrt{48\pi}.
\end{eqnarray}
The end of the slow roll occurs
when either or both of these inequalities
are saturated, at a value of $\phi$ denoted by $\phi_{\rm end}$.
Since $H\equiv {\dot R} /R$,
or $Hdt = d\ln R$, it follows that
\begin{equation}
d\ln R = {8\pi\over \mpl^2}\, {V(\phi ) d\phi
\over -V^\prime (\phi )} = -{8\pi d\phi \over \mpl \,x}.
\end{equation}
Now express the cosmic-scale factor
in terms of is value at the end of inflation, $R_{\rm end}$,
and the number of e-foldings before the end of inflation, $N(\phi )$,
$$R = \exp [-N(\phi )]\,R_{\rm end}. $$
The quantity $N(\phi )$ is a time-like variable whose
value at the end of inflation is zero and whose
evolution is governed by
\begin{equation}\label{eq:neq}
{dN \over d\phi} = {8\pi \over \mpl\,x} .
\end{equation}
Using Eq. (\ref{eq:neq}) we can compute the value of
the scalar field 50 e-folds before the end of inflation ($\equiv
\phi_{50}$); the values of
$V_{50}$, $x_{50}$, and $x_{50}^\prime$ follow directly.

As $\phi$ rolls down its potential during inflation its
energy density decreases,
and so the growth in the scale factor is not exponential.
By using the fact that the stress-energy of the scalar
field takes the perfect-fluid form, we can solve
for evolution of the cosmic-scale factor.  Recall,
for the equation of state $p=\gamma \rho$, the scale factor grows as
$R\propto t^q$, where $q= 2/3(1+\gamma )$.  Here,
\begin{eqnarray}\label{eq:eos}
\gamma & = & {{1\over 2}{\dot\phi}^2 - V \over
        {1\over 2}{\dot\phi}^2 +V} = { x^2-48\pi \over
        x^2 +48\pi };  \\
q & = & {1\over 3} + {16\pi \over x^2}.
\end{eqnarray}
Since the steepness of the potential can change during
inflation, $\gamma$ is not in general constant;
the power-law index $q$ is more
precisely the logarithmic rate of the change of the
logarithm of the scale factor, $q=d\ln R/d\ln t$.

When the steepness parameter is small,
corresponding to a very flat potential, $\gamma$ is close
to $-1$ and the scale factor grows as a very large
power of time.  To solve the horizon problem the
expansion must be ``superluminal'' (${\ddot R}>0$),
corresponding to $q>1$, which requires that $x^2< 24\pi$.
Since ${1\over 2}{\dot\phi}^2/V = x^2/48\pi$,
this implies that ${1\over 2}{\dot\phi}^2 /V(\phi ) < {1\over 2}$,
justifying neglect of the scalar-field kinetic energy
in computing the expansion rate for all but the
steepest potentials.  (In fact there are much stronger
constraints; the COBE DMR data imply that $n \ga 0.5$,
which restricts $x_{50}^2 \la 4\pi$, ${1\over 2}{\dot\phi}^2/V
\la {1\over 12}$, and $q\ga 4$.)

Next, let us relate the size of a given scale to
when that scale crosses outside
the Hubble radius during inflation, specified by $N_1(\lambda )$,
the number of e-folds before the end of inflation.
The physical size of a perturbation is related to its comoving size,
$\lambda_{\rm phys}=R\lambda$; with the usual
convention, $R_{\rm today} =1$, the comoving size
is the physical size today.  When the scale $\lambda$
crosses outside the Hubble radius $R_1
\lambda = H_1^{-1}$.  We then assume that:  (1) at the end of
inflation the energy density is ${\cal M}^4\simeq
V(\phi_{\rm end})$; (2) inflation is followed by a
period where the energy density of the Universe is dominated by coherent
scalar-field oscillations which decrease as
$R^{-3}$; and (3) when value of the scale factor
is $R_{\rm RH}$ the Universe reheats to a temperature
$T_{\rm RH} \simeq \sqrt{\mpl \Gamma}$ and expands
adiabatically thereafter.  The ``matching equation''
that relates $\lambda$ and $N_1(\lambda )$ is:
\begin{equation}\label{eq:prematch}
\lambda =  {R_{\rm today}\over R_1 } H_1^{-1}
        = {R_{\rm today} \over R_{\rm RH}} \,
        {R_{\rm RH}\over R_{\rm end}}\,
        {R_{\rm end}\over R_1}\, H_1^{-1}.
\end{equation}
Adiabatic expansion since reheating implies
$R_{\rm today}/R_{\rm RH} \simeq T_{\rm RH}/2.73\,{\rm K}$;
and the decay of the coherent scalar-field oscillations
implies $(R_{\rm RH}/R_{\rm end})^3 = ({\cal M}/T_{\rm RH})^4$.
If we define ${\bar q} = \ln (R_{\rm end}/R_1)/\ln (t_{\rm end}/t_1)$,
the mean power-law index, it follows that $(R_{\rm end}/R_1)H_1^{-1} =
\exp [N_1({\bar q}-1)/{\bar q}]H_{\rm end}^{-1}$,
and Eq. (\ref{eq:prematch}) becomes
\begin{equation}\label{eq:match}
N_1(\lambda )=  {{\bar q}\over {\bar q} -1}\,
\left[48+\ln \lambda_{\Mpc}
+ {4\over 3}\ln({\cal M}/10^{14}\GeV) - {1\over 3}
\ln (T_{\rm RH}/10^{14}\GeV ) \right];
\end{equation}
In the case of perfect reheating, which probably only applies to
first-order inflation, $T_{\rm RH} \simeq{\cal M}$.

The scales of astrophysical interest
today range roughly from that of galaxy size,
$\lambda \sim \Mpc$, to the present
Hubble scale, $H_0^{-1}\sim 10^4\Mpc$; up to the
logarithmic corrections these
scales crossed outside the horizon between about
$N_1(\lambda )\sim 48$ and $N_1(\lambda )\simeq 56$
e-folds before the end of inflation.  {\it That is,
the interval of inflation that determines its all observable
consequences covers only about 8 e-folds.}

Except in the case of strict power-law inflation
$q$ varies during inflation; this means that the
$(R_{\rm end}/R_1)H_1^{-1}$ factor in Eq. (\ref{eq:prematch})
cannot be written in closed form.
Taking account of this, the matching equation becomes
a differential equation,
\begin{equation}\label{eq:dmatch}
{d\ln\lambda_{\Mpc}\over dN_1} = {q(N_1) -1  \over q(N_1)};
\end{equation}
subject to the ``boundary condition:''  $\ln \lambda_{\Mpc}
= -48 -{4\over 3}\ln ({\cal M}/10^{14}\GeV )+{1\over 3}\ln
(T_{\rm RH}/10^{14}\GeV )$ for $N_1=0$, the matching relation
for the mode that crossed outside the Hubble radius at the
end of inflation.  Equation (\ref{eq:dmatch}) allows
one to obtain the precise expression for when a given scale
crossed outside the Hubble radius during inflation.  To
actually solve this equation, one would need to supplement it
with the expressions $dN/d\phi = 8\pi /\mpl x$ and
$q = 16\pi /x^2$.  For our purposes we need only know:  (1) The scales of
astrophysical interest correspond to $N_1\sim ``50\pm 4$,''
where for definiteness we will throughout take this
to be an equality sign.  (2) The expansion of Eq.
(\ref{eq:dmatch}) about $N_1 =50$,
\begin{equation}\label{eq:dndl}
\Delta N_1(\lambda ) = \left( {q_{50}-1\over q_{50} } \right)
\Delta \ln \lambda_{\Mpc} ;
\end{equation}
which, with the aid of Eq. (\ref{eq:neq}), implies that
\begin{equation} \label{eq:philambda}
\Delta \phi = \left( {q_{50} -1\over q_{50}}\right)\,
{x_{50}\over 8\pi}\, \Delta \lambda_{\Mpc} .
\end{equation}

We are now ready to express the perturbations
in terms of $V_{50}$, $x_{50}$, and
$x_{50}^\prime$.  First, we must solve for the value of $\phi$,
50 e-folds before the end of inflation.
To do so we use Eq. (\ref{eq:neq}),
\begin{equation}\label{eq:phi50}
N(\phi_{50} ) = 50 = {8\pi \over \mpl^2}\int_{\phi_{\rm end}}^{\phi_{50}}
{Vd\phi \over V^\prime}.
\end{equation}
Next, with the help of Eq. (\ref{eq:philambda})
we expand the potential $V$ and its steepness
$x$ about $\phi_{50}$:
\begin{equation}\label{eq:expandV}
V \simeq V_{50} + V_{50}^\prime (\phi - \phi_{50} )
= V_{50}\left[ 1 +  {x_{50}^2\over 8\pi}\,\left( {q_{50}\over q_{50}-1}
\right) \,  \Delta\ln \lambda_{\Mpc} \right] ;
\end{equation}
\begin{equation}\label{eq:expandx}
x \simeq x_{50} + x^\prime_{50}(\phi - \phi_{50})
= x_{50}\left[ 1 + {\mpl x_{50}^{\prime}
\over 8\pi}\,\left( {q_{50} \over q_{50}-1}\right)
\,\Delta\ln\lambda_{\Mpc} \right];
\end{equation}
of course these expansions only make sense for
potentials that are smooth.  We note that additional
terms in either expansion are ${\cal O}(\alpha_i^2)$ and
beyond the accuracy we are seeking.

Now recall the equations for the amplitude of the
scalar and tensor perturbations,
\begin{eqnarray}\label{eq:perts}
(\delta \rho /\rho )_{{\rm HOR},\lambda} & = & c_S \left( {V^{1/2}\over
\mpl^2 x}\right)_1 ;\\
h_{{\rm HOR},\lambda} & = & c_T\left( {V^{1/2}\over \mpl^2} \right)_1 ;
\end{eqnarray}
where subscript 1 means that the quantities are to be
evaluated where the scale $\lambda$ crossed outside the
Hubble radius, $N_1(\lambda )$ e-folds before the
end of inflation.   The origin of any deviation from
scale invariance is clear:  For tensor perturbations it
arises due to the variation of the potential; and for
scalar perturbations it arises due to the
variation of both the potential and its steepness.

Using Eqs. (\ref{eq:dndl}-22) it is now
simple to calculate the power-law exponents $\alpha_S$
and $\alpha_T$ that quantify the deviations from scale
invariance,
\begin{eqnarray}\label{eq:alpha}
\alpha_T  &  =  & {x_{50}^2\over 16\pi}\,{q_{50}\over q_{50}-1} \simeq
        {x_{50}^2 \over 16\pi};  \\
\alpha_S &  =  &  \alpha_T - {\mpl x_{50}^\prime \over 8\pi}\,{q_{50}
\over q_{50} -1} \simeq {x_{50}^2\over 16\pi}
- {\mpl x_{50}^\prime \over 8\pi};
\end{eqnarray}
where
\begin{eqnarray}\label{eq:defofsi}
q_{50} &  =  &  {1\over 3} + {16\pi \over x_{50}^2}
        \simeq {16\pi \over x_{50}^2} ;\\
h_{{\rm HOR},\lambda} & = & c_T \left( {V_{50}^{1/2}\over
\mpl^2 } \right) \, \lambda_{\Mpc}^{\ \ \,\alpha_T};\\
(\delta\rho /\rho )_{{\rm HOR},\lambda} &  =  &  c_S
\left({V_{50}^{1/2}\over x_{50}\mpl^2}\right)\,
\lambda_{\Mpc}^{\ \ \,\alpha_S}.
\end{eqnarray}
Note that the deviations from scale invariance, quantified
by $\alpha_S$ and $\alpha_T$, are of the order of
$x_{50}^2$, $\mpl x_{50}^\prime$.  In deriving the
expressions above we retained only the lowest-order
contributions; in all expressions there are
higher-order terms, ${\cal O}(\alpha_i^2) \sim
{\cal O}[x_{50}^2,  x_{50}(\mpl x_{50}^\prime ),
(\mpl x_{50}^\prime )^2]$.  The corrections to the
spectral indices are ${\cal O}(\alpha_i^2)$, and those
to the amplitudes are ${\cal O}(\alpha_i)$.
The justification for truncating the expansion at
lowest order is that the deviations from scale invariance
are expected to be small.

As we discuss in more detail in the Appendix, our more
intuitive power-law indices $\alpha_S$, $\alpha_T$ are
related to the indices that are usually used
to describe the power spectra of scalar and tensor
perturbations, $P_S(k) = |\delta_k|^2 = A k^n$
and $P_T(k) = |h_k|^2 = A_Tk^{n_T}$,
\begin{eqnarray}\label{eq:indices}
n & = & 1-2\alpha_S = 1 -{x_{50}^2\over 8\pi}
+ {\mpl x_{50}^\prime\over 4\pi} ; \\
n_T & = & -2\alpha_T =  -{x_{50}^2\over 8\pi} .\\
\end{eqnarray}

Finally, let us be more specific about the amplitude of the
scalar and tensor perturbations; in particular, for small $\alpha_S$,
$\alpha_T$ the contributions of each to
the quadrupole CBR temperature anisotropy:
\begin{eqnarray}\label{eq:quadanisotropy}
\left({\Delta T\over T_0}\right)_{Q-S}^2  &  \approx  &
{32\pi\over 45}{V_{50}\over \mpl^4 x_{50}^2};\\
\left({\Delta T\over T_0}\right)_{Q-T}^2  & \approx  &
0.61{V_{50}\over \mpl^4};
\end{eqnarray}
\begin{equation}\label{eq:ratio}
{T\over S}  \equiv {(\Delta T/T_0)_{Q-T}^2 \over
(\Delta T/T_0)_{Q-S}^2} \approx {0.28  x_{50}^2};
\end{equation}
where as usual expressions have been evaluated
to lowest order in $x_{50}^2$ and $\mpl x_{50}^\prime$.
These quantities represent the ensemble averages of the
scalar and tensor contributions to the
quadrupole temperature anisotropy, which in terms of
the spherical-harmonic expansion of the CBR temperature anisotropy
on the sky are given by $5\langle |a_{2m}|^2\rangle /4\pi$.  Further,
the scalar and tensor contributions
to the {\it measured} quadrupole anisotropy add
in quadrature, and are subject
to ``cosmic variance.''   (Cosmic
variance refers to the dispersion in the values measured
by different observers in the Universe.)  We refer the reader to the
Appendix for more details.

Before going on to specific models, let us make some general
remarks.  The steepness parameter $x_{50}^2$ must
be less than about $24\pi$ to ensure
superluminal expansion.   For ``steep'' potentials,
the expansion rate is ``slow,'' i.e., $q_{50}$ closer to unity,
the gravity-wave contribution to the quadrupole CBR temperature anisotropy
becomes comparable to, or greater than, that of density
perturbations, and both scalar and tensor
perturbations exhibit significant deviations from scale invariance.
For ``flat'' potentials, i.e., small $x_{50}$,
the expansion rate is ``fast,'' i.e., $q_{50} \gg 1$,
the gravity-wave contribution to the quadrupole CBR temperature anisotropy
is much smaller than that of density perturbations, and
the tensor perturbations are scale invariant.  Unless
the steepness of the potential changes rapidly, i.e.,
large $x_{50}^\prime$, the scalar perturbations are also
scale invariant.

\section {Worked Examples}

In this Section we apply the formalism developed
in the previous Section to four specific models.
So that we can, where appropriate, solve numerically
for model parameters, we will:  (1) Assume that
the astrophysically interesting scales crossed
outside the horizon 50 e-folds before the end
of inflation; and (2) Use the COBE DMR quadrupole measurement,
$\langle (\Delta T )_{Q-S}^2 \rangle^{1/2}
\approx 16\mu$K \cite{dmr}, to normalize
the scalar perturbations; using Eq. (\ref{eq:quadanisotropy}) this implies
\begin{equation}
V_{50} \approx 1.6\times 10^{-11} \,\mpl^4\,x_{50}^2 .
\end{equation}
We remind the reader that it is entirely possible that
a significant portion of the quadrupole anisotropy is
due to tensor-mode perturbations.  It is, of course, straightforward
to change ``50'' to the number appropriate to a
specific model, or to normalize the perturbations another way.

\subsection {Exponential potentials}

There are a class of models that can be described in terms
of an exponential potential,
\begin{equation}
V(\phi ) = V_0\exp (-\beta\phi /\mpl ).
\end{equation}
This type of potential was first invoked in the
context of power-law inflation \cite{powerlaw}, and has
recently received renewed interest in the context
of extended inflation \cite{extended}.  In the simplest
model of extended, or first-order, inflation,
that based upon the Brans-Dicke-Jordan theory
of gravity \cite{firstorder}, $\beta$ is related to the Brans-Dicke
parameter:  $\beta^2 = 64\pi /(2\omega +3)$.

For such a potential the slow-roll conditions are
satisfied provided that $\beta^2 \la 24\pi$;
thus inflation does not end until
the potential changes shape, or in the case of
extended inflation, until the phase transition
takes place.  In either case we can relate
$\phi_{50}$ to $\phi_{\rm end}$,
\begin{equation}
N(\phi_{50} ) = 50 = {8\pi\over \mpl^2}\int_{\phi_{50}}^{\phi_{\rm end}}
{Vd\phi \over -V^\prime};\qquad \Rightarrow\ \
\phi_{50} = \phi_{\rm end} - 50\beta / 8\pi .
\end{equation}
Since $\phi_{\rm end}$ is in effect arbitrary,
the overall normalization of the potential is
irrelevant.  The two other parameters,
$x_{50}$ and $x_{50}^\prime$, are easy to compute:
\begin{equation}
x_{50} = -\beta ; \qquad x_{50}^\prime = 0.
\end{equation}
Using the COBE DMR normalization, we can relate
$V_{50}$ and $\beta$:
\begin{equation}
V_{50} = 1.6 \times 10^{-11}\,\mpl^4 \beta^2.
\end{equation}
Further, we can compute $q$, $\alpha_S$, $\alpha_T$,
and $T/S$:
\begin{equation}
q = 16\pi /\beta^2; \qquad T/S = 0.28\beta^2;\qquad
\alpha_T =\alpha_S = 1/(q-1)\simeq \beta^2/16\pi .
\end{equation}
Note, for the exponential potential, $q$, $\alpha_T=
\alpha_S$ are independent of epoch.  In the case of
extended inflation, $\alpha_S =\alpha_T =4/(2\omega +3)$;
since $\omega$ must be less than about 20 \cite{erick},
this implies significant tilt:  $\alpha_S=\alpha_T \ga 0.1$.

\subsection{Chaotic inflation}

These models are based upon a very simple potential:
\begin{equation}
V(\phi ) = a\phi^b;
\end{equation}
$b=4$ corresponds to Linde's original model
of chaotic inflation and $a$ is dimensionless \cite{chaotic},
and $b=2$ is a model based upon a massive scalar field
and $m^2 = 2a$ \cite{massive}.  In these models $\phi$ is
initially displaced from $\phi = 0$, and inflation
occurs as $\phi$ slowly rolls to the origin.
The value of $\phi_{\rm end}$
is easily found:  $\phi_{\rm end}^2 = b(b-1)\mpl^2/24\pi$, and
\begin{eqnarray}
N(\phi_{50})=50 & =  & {8\pi \over \mpl^2}\int_{\phi_{\rm end}}^{\phi_{50}}
{Vd\phi \over V^\prime} ;\\
& \Rightarrow & \ \
\phi_{50}^2/\mpl^2  =  50b/4\pi + b^2/48\pi \simeq 50b/4\pi ;
\end{eqnarray}
the value of $\phi_{50}$ is a few times the Planck mass.

For purposes of illustration consider $b=4$; $\phi_{\rm end}
=\mpl /\sqrt{2\pi} \simeq 0.4\mpl$, $\phi_{50} \simeq
4\mpl$, $\phi_{46} \simeq 3.84\mpl$,
and $\phi_{54}\simeq 4.16\mpl$.  In order to have sufficient
inflation the initial value of $\phi$ must exceed about
$4.2\mpl$; inflation ends when $\phi \approx 0.4\mpl$; and
the scales of astrophysical interest cross outside the horizon
over an interval $\Delta \phi \simeq 0.3\mpl$.

The values of the potential, its steepness, and the
change in steepness are easily found,
\begin{equation}
V_{50} = a\,\mpl^b\,\left({50b\over 4\pi}\right)^{b/2};
\qquad x_{50} = \sqrt{4\pi b\over 50}; \qquad
\mpl x_{50}^\prime = {-4\pi \over 50} ;
\end{equation}
\begin{equation}
q_{50} = 200/b; \qquad
T/S = 0.07b; \qquad \alpha_T\simeq b/200; \qquad
\alpha_S = \alpha_T + 0.01 .
\end{equation}
Unless $b$ is very large, scalar perturbations dominate
tensor perturbations \cite{star}, $\alpha_T$, $\alpha_S$ are very
small, and $q$ is very large.  Further, when $\alpha_T$,
$\alpha_S$ become significant, they are equal.
Using the COBE DMR normalization we find:
\begin{equation}
a = 1.6\times 10^{-11} b^{1-b/2} (4\pi /50)^{b/2+1}\,\mpl^{4-b}.
\end{equation}
For the two special cases of interest: $b=4$, $a=6.4\times 10^{-14}$;
and $b=2$, $m^2 \equiv 2a = 2.0\times 10^{-12}\mpl^2$.

\subsection {New inflation}

These models entail
a very flat potential where the scalar field rolls from
$\phi \approx 0$ to the minimum of the potential at $\phi
=\sigma$.  The original models of slow-rollover inflation \cite{new}
were based upon potentials of the Coleman-Weinberg form
\begin{equation}
V(\phi ) = B\sigma^4/2 +B\phi^4\left[ \ln (\phi^2/\sigma^2)
-{1\over 2} \right] ;
\end{equation}
where $B$ is a very small dimensionless
coupling constant.  Other very flat potentials also work (e.g.,
$V = V_0 - \alpha\phi^4 +\beta \phi^6$ \cite{st}).  As before
we first solve for $\phi_{50}$:
\begin{equation}
N(\phi_{50}) = 50 = {8\pi\over \mpl^2}\int_{\phi_{\rm end}}^{\phi_{50}}
{Vd\phi\over V^\prime};\qquad \Rightarrow \ \
\phi_{50}^2 = {\pi \sigma^4\over 100 |\ln (\phi_{50}^2/\sigma^2)|\mpl^2};
\end{equation}
where the precise value of $\phi_{\rm end}$ is not relevant,
only the fact that it is much larger than $\phi_{50}$.
Provided that $\sigma \la \mpl$, both $\phi_{50}$ and
$\phi_{\rm end}$ are much less than $\sigma$; we then find
\begin{equation}
V_{50}\simeq B\sigma^4/2; \qquad x_{50} \simeq - {(\pi /25)^{3/2}
\over \sqrt{|\ln (\phi_{50}^2/\sigma^2|) }} \left( {\sigma\over \mpl}
\right)^2 \ll 1;
\end{equation}
\begin{equation}
\mpl x_{50}^\prime \simeq -24\pi /100;\qquad
q_{50} \simeq {2.5\times 10^5 |\ln (\phi_{50}^2/\sigma^2)|\over \pi^2}
\left({\mpl\over \sigma}\right)^4 \gg 1;
\end{equation}
\begin{equation}
\alpha_S \simeq {1\over q_{50}} \ll 1;\qquad
\alpha_T = \alpha_S + 0.03;\qquad
{T\over S} \simeq {6\times 10^{-4}\over |\ln (\phi_{50}^2/\sigma^2)|}\left(
{\sigma\over \mpl}\right)^4.
\end{equation}
Provided that $\sigma \la \mpl$, $x_{50}$ is very small,
implying that $q$ is very large, the gravitational-wave
and density perturbations are very nearly scale invariant,
and $T/S$ is small.  Finally, using the COBE DMR normalization,
we can determine the dimensionless coupling constant $B$:
\begin{equation}
B \simeq 6\times 10^{-14}/|\ln (\phi_{50}^2/\sigma^2)|
\approx 3\times 10^{-15}.
\end{equation}

\subsection{Natural inflation}

This model is based upon a potential of the form \cite{freese}
\begin{equation}
V(\phi ) = \Lambda^4 \left[ 1+\cos (\phi /f ) \right].
\end{equation}
The flatness of the potential (and requisite small
couplings) arise because the $\phi$ particle is a pseudo-Nambu-Goldstone
boson ($f$ is the scale of spontaneous symmetry breaking
and $\Lambda$ is the scale of explicit symmetry breaking; in
the limit that $\Lambda \rightarrow 0$ the $\phi$ particle
is a massless Nambu-Goldstone boson).   It is a simple
matter to show that $\phi_{\rm end}$ is of the order of $\pi f$.

This potential is difficult to analyze in
general; however, there are two limiting regimes:
(i) $f\gg \mpl$; and (ii) $f\la \mpl$ \cite{st}.  In the first
regime, the 50 or so relevant e-folds take place close
to the minimum of the potential, $\sigma = \pi f$, and
inflation can be analyzed by expanding the potential about $\phi=\sigma$,
\begin{equation}
V(\psi ) \simeq {m^2\psi^2}/2 ;
\end{equation}
\begin{equation}
m^2 = \Lambda^4 /f^2; \qquad \psi = \phi -\sigma .
\end{equation}
In this regime natural inflation is equivalent to chaotic
inflation with $m^2 =\Lambda^4 /f^2 \simeq 2\times 10^{-12}\mpl^2$.

In the second regime, $f\la \mpl$, inflation takes
place when $\phi \la \pi f$, so that we can make the
following approximations:  $V \simeq 2\Lambda^4$
and $V^\prime = - \Lambda^4\phi /f^2$.  Taking
$\phi_{\rm end} \sim \pi f$, we can solve for $N(\phi )$:
\begin{equation}
N(\phi ) = {8\pi \over \mpl^2}\int_\phi^{\pi f} {Vd\phi \over -V^\prime}
\simeq {16\pi \mpl^2\over f^2} \ln (\pi f/\phi ) ;
\end{equation}
from which it is clear that achieving 50 e-folds of
inflation places a lower bound to $f$, very
roughly $f\ga \mpl /3$ \cite{st,freese}.

Now we can solve for $\phi_{50}$, $V_{50}$, $x_{50}$,
and $x_{50}^\prime$:
\begin{equation}
\phi_{50}/\pi f \simeq \exp (-50 \mpl^2/16\pi f^2)\la {\cal O}(0.1);
\qquad V_{50} \simeq 2\Lambda^4;
\end{equation}
\begin{equation}
x_{50} \simeq {1\over 2}\,{\mpl\over f}\,{\phi_{50}\over f}
\la {\cal O}(0.1) ;\qquad
x_{50}^\prime \simeq - {1\over 2}\,\left( {\mpl\over f}\right)^2.
\end{equation}
Using the COBE DMR normalization, we can
relate $\Lambda$ to $f/\mpl$:
\begin{equation}
\Lambda /\mpl = 6.7\times 10^{-4} \sqrt{\mpl\over f}
\exp (-25\mpl^2/16\pi f^2).
\end{equation}
Further, we can solve for $T/S$, $\alpha_T$, and $\alpha_S$:
\begin{equation}
{T\over S} \simeq 0.07 \left( {\mpl\over f}\right)^2
\left({\phi_{50} \over f}\right)^2 \la {\cal O}(0.1) ;
\end{equation}
\begin{equation}
\alpha_T = {1\over 16\pi}\,{q_{50}\over q_{50}-1} \left(
{1\over 4} {\mpl^2\over f^2}{\phi_{50}^2\over f^2} \right)
\approx  {1\over 64\pi}
\left({\mpl\over f}\right)^2\left({\phi_{50}\over f}\right)^2\ll 0.1;
\end{equation}
\begin{equation}
\alpha_S = {1\over 16\pi}\,{q_{50}\over q_{50}-1} \left(
{1\over 4}{\mpl^2\over f^2}{\phi_{50}^2\over f^2} + {\mpl^2\over
f^2}\right) \approx {1\over 16\pi}\left({\mpl\over f}\right)^2;
\end{equation}
\begin{equation}
q_{50} = 64\pi \left({f\over \mpl}\right)^2\left(
{f\over \phi_{50}}\right)^2  \gg 1.
\end{equation}

Regime (ii) provides the exception
to the rule that $\alpha_S\approx\alpha_T$ and large
$\alpha_S$ implies large $T/S$.  For example, taking
$f=\mpl /2$, we find:
\begin{equation}
\phi_{50}/f \sim 0.06; \qquad x_{50} \sim 0.06; \qquad
x_{50}^\prime = - 2; \qquad q_{50} \sim 10^4;
\end{equation}
\begin{equation}
\alpha_T \sim 10^{-4};\qquad \alpha_S \sim 0.08;\qquad T/S \sim 10^{-3}.
\end{equation}
The gravitational-wave perturbations are very nearly scale
invariant, while the density perturbations deviate
significantly from scale invariance.  We note that this
regime (ii), i.e., $f \la \mpl$, occupies only a tiny fraction of parameter
space because $f$ must
be greater than about $\mpl /3$ to achieve sufficient
inflation; further, regime (ii) is ``fine tuned'' and
``unnatural'' in the sense that the required value of $\Lambda$ is
exponentially sensitive to the value of $f/\mpl$.

Finally, we note that the results for regime (ii)
apply to any inflationary model whose Taylor expansion
in the inflationary region is similar; e.g., $V(\phi )=
-m^2\phi^2 + \lambda\phi^4$, which was originally analyzed
in Ref.~\cite{st}.

\section{Concluding Remarks}

Beyond the generic prediction of a flat Universe and its
important consequences for the matter content of the
Universe, namely that most of the matter in the
Universe is nonbaryonic \cite{dark},
the observable consequences of
inflation are tied to density and gravity-wave
perturbations.  (In models of first-order inflation
vacuum-bubble collisions provide a very potent
source of short-wavelength gravity waves \cite{gws}.)
The amplitude and spectrum of these perturbations depend
upon the shape of the inflationary potential in the narrow
interval where the scalar field was
around ``$50\pm 4$'' e-folds before the end of inflation.  By expanding
the potential about this interval in terms of its
value, $V_{50}$, its steepness, $x_{50} = [\mpl V^\prime /V]_{50}$,
and the rate of change of its steepness, $x_{50}^\prime$,
we have expressed the amplitudes and power-law indices
of the scalar and tensor metric perturbations in terms
of these three quantities to lowest order in the
deviations from scale invariance.   Measurements of the
amplitudes and spectral indices of the density
and gravity-wave perturbations determine---in fact
over determine---$V_{50}$, $x_{50}$,
and $x_{50}^\prime$, and, in principle, such measurements
allow one to both infer the shape of the
inflationary potential and to test the consistency of the
inflationary hypothesis \cite{reconstruct}.

There are limitations to our formalism; it
is not applicable to potentials that are not ``smooth'' or
have inclines in the region that affects
astrophysically interesting scales.
This includes potentials with ``specially engineered''
bumps and wiggles \cite{designer}.

To summarize the general features of our results.
In all examples the deviations from scale invariance
enhance perturbations on large scales.  The only
potentials that have significant deviations from
scale invariance are very steep or have rapidly
changing steepness.  In the former case, both the
scalar and tensor perturbations are tilted by a
similar amount; in the latter case, only the scalar
perturbations are tilted.

For ``steep'' potentials,
the expansion rate is ``slow,'' i.e., $q_{50}$ close to unity,
the gravity-wave contribution to the CBR quadrupole anisotropy
becomes comparable to, or greater than, that of density
perturbations, and both scalar and tensor
perturbations are tilted significantly.
For flat potentials, i.e., small $x_{50}$,
the expansion rate is ``fast,'' i.e., $q_{50} \gg 1$,
the gravity-wave contribution to the CBR quadrupole
is much smaller than that of density perturbations, and unless
the steepness of the potential changes significantly,
large $x_{50}^\prime$, both spectra very nearly scale invariant;
if the steepness of the potential changes rapidly,
the spectrum of scalar perturbations can be tilted significantly.
The models that permit significant deviations from scale
invariance involve exponential or cosine potentials.
The former by virtue of their steepness, the latter
by virtue of the rapid variation of their steepness.

Only recently has the deviation of the metric perturbations
from the scale-invariant Harrison-Zel'dovich form drawn
intense scrutiny, though their deviation from scale
invariance has been noted since the very beginning
\cite{scalar,st}.  This new interest traces in part to
the growing body of observational data that are putting
the cold dark matter scenario to the test:
The COBE DMR result, together with
a host of other observations, {\it may} be inconsistent
with the simplest version of cold dark matter, that with
scale-invariant density perturbations ($\alpha_S=0$).
(Then again, the problems may disappear.)
A slight deviation from scale
invariance or tilt, $\alpha_S \simeq 0.08$ or $n\approx 0.84$,
seems to improve concordance with the observational data by reducing
the amplitude of scalar perturbations on small
scales \cite{tilt}.  Of the models analyzed here,
only two permit significant tilt,
those based on exponential potentials, which
include the very attractive extended-inflation models,
and natural inflation.
The former are also characterized by significant
tensor contribution to the quadrupole anisotropy, while
the latter are not; a separation of the tensor and
scalar contributions could cleanly distinguish
between these two types of models.  And in that regard,
measurements of CBR anisotropy on angular scales of less than a
few degrees will play a crucial role.

\bigskip
\bigskip

\noindent I wish to thank Paul Steinhardt and James Lidsey for
valuable conversations.  This work was supported
in part by the DOE and by the NASA through NAGW-2381 (at Fermilab).

\bigskip
\bigskip
\bigskip

\appendix
\section{Appendix}

In Section II we were purposefully vague when discussing
the amplitudes of the scalar and tensor modes, except
when specifying their contributions to the quadrupole CBR temperature
anisotropy; in fact, the spectral indices $\alpha_S$ and
$\alpha_T$, together with the scalar and tensor
contributions to the CBR quadrupole serve to
provide all the information necessary.  In this
Appendix we fill in more of the details about the
metric perturbations.

The scalar and tensor metric perturbations are expanded
in harmonic functions, in the flat Universe predicted
by inflation, plane waves,
\begin{eqnarray}
h ({\bf x}, t) & = & {1\over (2\pi )^3}
\int d^3k\,h_{\bf k}^i (t)\, \varepsilon_{\mu\nu}^i
\, e^{-i{\bf k}\cdot{\bf x}} ;\\
{\delta\rho ({\bf x},t) \over \rho} & = &
{1\over (2\pi )^3}\int d^3k\,\delta_{\bf k} (t) \, e^{-i{\bf k}\cdot{\bf x}} ;
\end{eqnarray}
where $\varepsilon_{\mu\nu}^i$ is the polarization tensor
for the gravity-wave modes, and $i= +$, $\times$ are
the two polarization states.  Everything of interest
can be computed in terms of $h_{\bf k}^i$ and $\delta_{\bf k}$.
For example, the {\it rms} mass fluctuation
in a sphere of radius $r$ is obtained in terms of the
window function for a sphere and the power spectrum $P_S(k)
\equiv \langle |\delta_{\bf k}|^2\rangle$ (see below),
\begin{equation}
\langle (\delta M /M)^2\rangle = {9\over 2\pi^2r^2}\,
\int_0^\infty [j_1(kr)]^2 \,P_S(k) dk ;
\end{equation}
$j_1(x)$ is the spherical Bessel function of first order.
Roughly, $(\delta M/M)^2 \sim k^3|\delta_{\bf k}|^2$,
evaluated on the scale $k=r^{-1}$.
This is what we meant by $(\delta \rho /
\rho)_{{\rm HOR},\lambda}$:  the {\it rms} mass
fluctuation on the scale $\lambda$
when it crossed inside the horizon.  Likewise,
by $h_{{\rm HOR},\lambda}$ we meant the {\it rms} strain
on the scale $\lambda$ as it crossed inside the Hubble radius,
$(h_{{\rm HOR},\lambda})^2 \sim k^3|h_{\bf k}^i|^2$.

In the previous discussions we have chosen to specify
the metric perturbations for the different Fourier
modes when they crossed inside the horizon,
rather than at a common time.  We did so because
scale invariance is made manifest, as the scale independence
of the metric perturbations at post-inflation horizon crossing.
Further, in the case of scalar perturbations
$(\delta \rho /\rho )_{\rm HOR}$ is up to a numerical factor
the fluctuation in the Newtonian potential, and, by specifying
the scalar perturbations at horizon crossing, we avoid the
discussion of scalar perturbations on superhorizon
scales, which is beset by the subtleties associated with
the gauge noninvariance of $\delta_{\bf k}$.

It is, however, necessary to specify the perturbations at a common time
to carry out most calculations; e.g., an $N$-body simulation
of structure formation or the calculation of CBR anisotropy.
To do so, one has to
take account of the evolution of the perturbations
after they enter the horizon.
After entering the horizon tensor perturbations behave like
gravitons, with $h_{\bf k}$ decreasing as $R^{-1}$ and
the energy density associated with a given mode, $\rho_k \sim
\mpl^2 k_{\rm phys}^2k^3|h_{\bf k}|^2$, decreasing
as $R^{-4}$.  The evolution
of scalar perturbations is slightly more complicated; modes that
enter the horizon while the Universe is still radiation dominated
remain essentially constant until the Universe becomes matter
dominated (growing only logarithmically);
modes that enter the horizon after the Universe becomes
matter dominated grow as the scale factor.
(The gauge noninvariance of $\delta_{\bf k}$ is not an important
issue for subhorizon size modes; here a Newtonian analysis
suffices, and there is only one growing mode, corresponding to
a density perturbation.)

The method for characterizing the scalar perturbations
is by now standard:  The spectrum of
perturbations is specified at the present
epoch (assuming linear growth for all scales); the spectrum at earlier
epochs can be obtained by multiplying $\delta_{\bf k}$
by $R(t)/R_{\rm today}$.  First, it should be noted
that $\delta_{\bf k}$ is a gaussian, random variable with
statistical expectation
\begin{equation}
\langle \delta_{\bf k}\,\delta_{\bf q}\rangle
= P_S(k) \delta^{(3)} ({\bf k} -{\bf q});
\end{equation}
where the power spectrum today is written as
\begin{equation}
P_S(k) \equiv Ak^n T(k)^2;
\end{equation}
$n=1-2\alpha_S$ ($=1$ for scale-invariant perturbations), and
$T(k)$ is the ``transfer function'' which encodes
the information about the post-horizon crossing evolution
of each mode and depends
upon the matter content of the Universe, e.g., baryons plus cold dark
matter, hot dark matter, warm dark matter, and so on \cite{stat}.
The transfer function is defined so that $T(k)\rightarrow 1$
for $k\rightarrow 0$ (long-wavelength perturbations); an
analytic approximation to the cold dark matter transfer
function is given by
\begin{equation}
T(k)   =  {\ln (1+2.34 q)/2.34q \over \left[ 1 + (3.89q) +(16.1q)^2
+ (5.46q)^3 + (6.71q)^4 \right]^{1/4}} ;
\end{equation}
where $q= k/(\Omega h^2 \Mpc^{-1})$.  The overall normalization factor
\begin{equation}
A = {1024\pi^3 \over 75H_0^4}\,{V_{50} \over \mpl^4 x_{50}^2}
\,{(1+x_{50}^2/16\pi )\over (k_{50}H_0/2)^{n-1} } ;
\end{equation}
where the ${\cal O}(\alpha_i)$ correction to
has been included and the quantity
$k_{50}$ is the comoving wavenumber of the scale
that crossed outside the horizon 50 e-folds before the
end of inflation.  All the formulas below simplify if this
scale corresponds to the present horizon scale, i.e., $k_{50}\sim H_0$.

{}From this expression it is simple to compute the
Sachs-Wolfe contribution of scalar perturbations
to the CBR temperature anisotropy; on angular scales much greater
than about $1^\circ$ (corresponding to multipoles $l \ll 100$) it
is the dominant contribution.  If we expand the
CBR temperature on the sky in spherical harmonics,
\begin{equation}
{\delta T(\theta ,\phi )\over T_0} = \sum_{l\ge 2, m=-l}^{l=\infty ,m=l}
a_{lm}Y_{lm}(\theta ,\phi );
\end{equation}
where $T_0=2.73\,$K is the CBR temperature today, then
the ensemble expectation for the multipole coefficients
is given by
\begin{eqnarray}
\langle |a_{lm}|^2\rangle & = & {H_0^4\over 2\pi}\,
\int_0^\infty k^{-2}\,P_S(k)\,|j_l(k r_0)|^2\,dk ; \\
&  \simeq &  {AH_0^{3+n}\, r_0^{1-n}\over 16}\,
{\Gamma (l+{1\over 2}n-{1\over 2})
\Gamma (3-n) \over \Gamma (l-{1\over 2}n +{5\over 2})
[\Gamma (2-{1\over 2}n)]^2} ;
\end{eqnarray}
where $r_0\approx 2H_0^{-1}$ is the comoving distance to the last scattering
surface, and this expression is for the Sachs-Wolfe contribution from scalar
perturbations only.  For $n$ not too different from
one, the ensemble expectation for the quadrupole CBR temperature
anisotropy is
\begin{equation}
\left( {\Delta T\over T_0} \right)_{Q-S}^2 \equiv
{5 |a_{2m}|^2 \over 4\pi } \approx {32\pi\over 45}\,
{V_{50} \over \mpl^4\,x_{50}^2}\,(k_{50}r_0)^{1-n}.
\end{equation}
(By choosing $k_{50}\sim r_0^{-1}= {1\over 2}H_0$,
the last factor becomes unity.)

The procedure for specifying the tensor modes is similar,
cf. Refs.~\cite{aw,white}.  For the modes that enter the
horizon after the Universe becomes matter dominated,
$k\la 0.1h^2\Mpc$, which are the only modes that contribute
significantly to CBR anisotropy on angular scales
greater than a degree,
\begin{equation}
h_{\bf k}^i (\tau ) = a^i ({\bf k}) \left( { 3j_1(k\tau )\over
k\tau }  \right) ;
\end{equation}
where $\tau = r_0(t/t_0)^{1/3}$ is conformal time.
[For the modes that enter the horizon during the radiation-dominated
era, $k \ga 0.1h^2\Mpc^{-1}$, the factor
$3j_1(k\tau )/k\tau$ is replaced by $j_0(k\tau )$
for the remainder of the radiation era.
In either case, the factor involving the spherical Bessel
function quantifies the fact that tensor perturbations
remain constant while outside the horizon, and after
horizon crossing decrease as $R^{-1}$.]

The tensor perturbations too are characterized by
a gaussian, random variable, here written as $a^i({\bf k})$;
the statistical expectation
\begin{equation}
\langle h_{\bf k}^i h_{\bf q}^j \rangle =
P_T (k) \delta^{(3)} ({\bf k} -{\bf q})\delta_{ij};
\end{equation}
where the power spectrum
\begin{eqnarray}
P_T(k) & = & A_T k^{n_T -3} \left[ {3j_1(k\tau )\over k\tau}\right]^2 ; \\
A_T & = & {8 \over 3\pi }\, {V_{50} \over \mpl^4}\,
{[\Gamma ({3\over 2} - {1\over 2}n_T)]^2 2^{-n_T}\over
[\Gamma ({3\over 2})]^2(1-{1\over 2}n_T)^2}\, k_{50}^{-n_T} ;
\end{eqnarray}
where the ${\cal O}(\alpha_i)$ correction has been
included.  Note that $n_T = -2\alpha_T$ is zero
for scale-invariant perturbations.

Finally, the contributions of the tensor perturbations to
the multipole amplitudes, which also arise due
to the Sachs-Wolfe effect \cite{sw,aw,white}, are given by
\begin{equation}
\langle |a_{lm}|^2 \rangle \simeq 36 \pi^2 \,{\Gamma (l+3)
\over \Gamma (l-1) }\, \int_0^\infty\,
k^{n_T+1}\,A_T \, |F_l(k)|^2\,dk ;
\end{equation}
where
\begin{eqnarray}
F_l(k) & =  & \int_0^{r_0} \, dr \,
\left[{d\over d(kr)}{j_1(kr)\over kr}\right]  \nonumber \\
& \times &  \left[ {j_{l-2}(kr) \over (2l-1)(2l+1)}
  + { 2j_l(kr)\over (2l-1)(2l+3)}
  + {j_{l+2}(kr)\over (2l+1)(2l+3)} \right] .
\end{eqnarray}
The tensor contribution to the quadrupole CBR temperature
anisotropy for $n_T$ not too different from zero is
\begin{equation}
\left( {\Delta T\over T_0} \right)_{Q-T}^2 \equiv
{5|a_{2m}|^2\over 4\pi} \simeq 0.61 {V_{50}\over \mpl^4}\, (k_{50}r_0)^{-n_T};
\end{equation}
where the integrals in the previous expressions have been evaluated
numerically.


\bigskip
\bigskip
\bigskip


\begin{thebibliography}  {inflation}

\bibitem{guth}  A.H.~Guth, {\it Phys. Rev. D} {\bf 23}, 347 (1981).

\bibitem{inflation} For a textbook discussion of inflation
see e.g., E.W.~Kolb and M.S.~Turner, {\it The Early
Universe} (Addison-Wesley, Redwood City, CA, 1990), Ch.~8.

\bibitem{scalar}  A.H.~Guth and S.-Y.~Pi, {\it Phys. Rev. Lett.}
{\bf 49}, 1110 (1982); A.A.~Starobinskii, {\it Phys. Lett. B}
{\bf 117}, 175 (1982); S.W.~Hawking, {\it ibid} {\bf 115}, 295 (1982);
J.M.~Bardeen, P.J.~Steinhardt, and M.S.~Turner, {\it Phys. Rev. D}
{\bf 28}, 679 (1983).

\bibitem{tensor} V.A.~Rubakov, M.~Sazhin, and A.~Veryaskin,
{\it Phys. Lett. B} {\bf 115}, 189 (1982); R.~Fabbri and
M.~Pollock, {\it ibid} {\bf 125}, 445 (1983); L.~Abbott
and M.~Wise, {\it Nucl. Phys. B} {\bf 244}, 541 (1984);
B.~Allen, {\it Phys. Rev. D} {\bf 37}, 2078 (1988).

\bibitem{hz} E.R.~Harrison, {\it Phys. Rev. D} {\bf 1},
2726 (1970); Ya.B.~Zel'dovich, {\it Mon. Not. R. astr. Soc.}
{\bf 160}, 1p (1972).

\bibitem{dmr} G.~Smoot et al., {\it Astrophys. J.} {\bf 396}, L1 (1992);
E.L.~Wright, {\it ibid} {\bf 396}, L3 (1992).

\bibitem{krauss} L.~Krauss and M.~White, {\it Phys. Rev. Lett.}
{\bf 69}, 869 (1992).

\bibitem{davisetal} R.~Davis et al., {\it Phys. Rev. Lett.}
{\bf 69}, 1856 (1992); F.~Lucchin, S.~Mattarese, and
S.~Mollerach, {\it Astrophys. J.} {\bf 401}, L49 (1992); D.~Salopek,
{\it Phys. Rev. Lett.} {\bf 69}, 3602 (1992); A.~Liddle and D.~Lyth,
{\it Phys. Lett. B} {\bf 291}, 391 (1992); J.E.~Lidsey and
P.~Coles, Queen Mary College preprint (1992); A.~Dolgov and J.~Silk,
unpublished (1992).

\bibitem{st} P.J.~Steinhardt and M.S.~Turner, {\it Phys. Rev. D}
{\bf 29}, 2162 (1984).

\bibitem{extenddavis} Some of what is presented here is a
comprehensive development and extension of ideas put
forth in R.~Davis et al., {\it Phys. Rev. Lett.}
{\bf 69}, 1856 (1992).

\bibitem{allslow}  At first sight, first-order inflation
might seem very different
from slow-rollover inflation, as reheating occurs through
the nucleation of percolation of true-vacuum bubbles.
However, such models can be recast as slow-rollover inflation by means
of a conformal transformation, and the analysis of
metric perturbations proceeds as in slow rollover inflation.
See e.g., E.W.~Kolb, D.~Salopek, and M.S.~Turner, {\it Phys.
Rev. D} {\bf 42}, 3925 (1990).

\bibitem{reheat} A.~Albrecht et al., {\it Phys. Rev. Lett.}
{\bf 48}, 1437 (1982); L.~Abbott and M.~Wise, {\it Phys. Lett. B}
{\bf 117}, 29 (1982); A.D.~Linde and A.~Dolgov, {\it ibid}
{\bf 116}, 329 (1982).

\bibitem{isocurv} See e.g., A.D.~Linde, {\it Phys. Lett. B}
{\bf 158}, 375 (1985); D.~Seckel and M.S.~Turner,
{\it Phys. Rev. D} {\bf 32}, 3178 (1985); M.S.~Turner,
A.~Cohen, and D.~Kaplan, {\it Phys. Lett. B} {\bf 216}, 20 (1989).

\bibitem{powerlaw} L.~Abbott and M.~Wise, {\it Nucl. Phys. B}
{\bf 244}, 541 (1984); F. Lucchin and S. Mattarese, {\it Phys. Rev. D}
{\bf 32}, 1316 (1985); R.~Fabbri, F.~Lucchin, and S.~Mattarese,
{\it Phys. Lett. B} {\bf 166}, 49 (1986).

\bibitem{extended}  D.~La and P.J.~Steinhardt,
{\it Phys. Rev. Lett.} {\bf 62}, 376 (1989).

\bibitem{firstorder} For a review of first-order inflation see e.g.,
E.W.~Kolb, {\it Physica Scripta} {\bf T36}, 199 (1991).

\bibitem{erick} E.~Weinberg, {\it Phys. Rev. D} {\bf 40}, 3950 (1989).

\bibitem{chaotic} A.D.~Linde, {\it Phys. Lett. B} {\bf 129}, 177 (1983).

\bibitem{massive}  V.~Belinsky, L.~Grishchuk, I.~Khalatanikov,
and Ya.B.~Zel'dovich, {\it Phys. Lett. B} {\bf 155}, 232
(1985); L.~Jensen, unpublished (1985).

\bibitem{star} A.A.~Starobinskii, {\it Sov. Astron.} {\bf 11}, 133 (1985).

\bibitem{new} A.D.~Linde, {\it Phys. Lett. B}
{\bf 108}, 389 (1982); A.~Albrecht and P.J.~Steinhardt,
{\it Phys. Rev. Lett.} {\bf 48}, 1220 (1982).

\bibitem{freese} K.~Freese, J.~Frieman, and
A.~Olinto, {\it Phys. Rev. Lett.} {\bf 65}, 3233 (1990).

\bibitem{dark}  See e.g., M.S.~Turner, {\it Physica Scripta}
{\bf T36}, 167 (1991); J.R.~Primack, B.~Sadoulet, and
D.~Seckel, {\it Ann. Rev. Nucl. Part. Sci.} {\bf 38}, 751 (1988).

\bibitem{gws}  M.S.~Turner and F.~Wilczek, {\it Phys.
Rev. Lett.} {\bf 65}, 3080 (1990);
A.~Kosowsky, Michael S.~Turner, and R.~Watkins
{\it ibid} {\bf 69}, 2026 (1992).

\bibitem{reconstruct} Attempts at the
reconstruction of the inflationary potential
from observational data include, H.M.~Hodges and
G.R.~Blumenthal, {\it Phys. Rev. D} {\bf 42}, 3329 (1990);
SCIPP preprint 89/56 (1989); and more recently,
E.~Copeland, E.W.~Kolb, A.~Liddle, and J.~Lidsey,
FERMILAB-Pub-93/0**-A (1993).

\bibitem{designer} D.~Salopek, J.R.~Bond, and J.M.~Bardeen,
{\it Phys. Rev. D} {\bf 40}, 1753 (1989); H.M.~Hodges and
G.R.~Blumenthal, {\it Phys. Rev. D} {\bf 42}, 3329 (1990).

\bibitem{tilt}  See e.g., J.P.~Ostriker, {\it Ann. Rev. Astron.
Astrophys.} {\bf 31}, in press (1993);
F.~Adams et al., {\it Phys. Rev. D} {\bf 47}, 426
(1993); J.~Gelb et al., {\it Astrophys. J.} {\bf 403},
L5 (1993); R.~Cen and J.P.~Ostriker, {\it ibid}, in press (1993).

\bibitem{stat} J.M.~Bardeen et al., {\it Astrophys. J.}
{\bf 304}, 15 (1986).

\bibitem{sw} R.K.~Sachs and A.M.~Wolfe, {\it Astrophys. J.}
{\bf 147}, 73 (1967).

\bibitem{aw} L.~Abbott and M.~Wise, {\it Nucl. Phys. B}
{\bf 244}, 541 (1984).

\bibitem{white} M.~White, {\it Phys. Rev. D} {\bf 46}, 4198 (1992).

\end{thebibliography}
\end{document}